# A how-to guide for code-sharing in biology

Richard J. Abdill[1*], Emma Talarico[2], Laura Grieneisen[2]

1 - Section of Genetic Medicine, Department of Medicine, University of Chicago, Chicago, Illinois, USA
2 - Department of Biology, University of British Columbia, Okanagan Campus, Kelowna, BC, Canada
* - correspondence to rabdill@uchicago.edu

# Abstract

Computational biology continues to spread into new fields, becoming more accessible to researchers trained in the wet lab who are eager to take advantage of growing datasets, falling costs, and novel assays that present new opportunities for discovery even outside of the much-discussed developments in artificial intelligence. However, guidance for implementing these techniques is much easier to find than guidance for reporting their use, leaving biologists to guess which details and files are relevant. Here, we provide a set of recommendations for sharing code, with an eye toward guiding those who are comparatively new to applying open science principles to their computational work. Additionally, we review existing literature on the topic, summarize the most common tips, and evaluate the code-sharing policies of the most influential journals in biology, which occasionally encourage code-sharing but seldom require it. Taken together, we provide a user manual for biologists who seek to follow code-sharing best practices but are unsure where to start.

# Introduction

Reproducible computational practices, and open science more broadly, are the subject of many discussions: transparency is good, but its implementation is time-intensive and poorly incentivized [1]. Open research is associated with more citations and more media coverage [2], but it can expose researchers to new avenues for harassment and suppression [3]. There are also many resources promoting particular approaches to performing computational work [4,5] and developing research software [6]. But amidst the discussions of how to perform computational research, where to publish it and how to organize your files, there is a dearth of information on how to be transparent about work that's already been done, particularly for biologists who may not specialize in computational work—many are unsure how to share code or, as a recent Springer Nature survey found, even where to upload it [7].

A complicating factor is the broad variation between fields in the standards for data and code sharing, as well as the types of data sets and code used. For example, data sharing standards are well-reviewed in the ecological literature [8,9], and major journals in the field of ecology have extensive data sharing policies such as Ecological Society of America's Open Research Policy 'Definitions' page [10]. However, these practices are less standardized in other fields of biology. This means that research and journals at the intersection of multiple fields, such as microbiome science's integration of medicine, ecology, and computational biology, may not have scientists trained in uniform standards of data and code sharing. Finally, many established best practices papers focus on how to design a study from the outset to fit into a reproducible science framework. However, science is messy, and as projects and datasets are passed between lab members, the researchers who are assembling the final paper for submission often inherit files and code that were not created using best practices. As such, a resource that guides researchers in how to apply open science sharing practices as best as possible to this 'inherited' code is needed.

The goal of our paper, therefore, is to provide an integrative guide for preparing and sharing code, such that these practices can be implemented across biological subfields and stages of the research process. We focus on the implementation details particular to *reproducible* results—those that can be regenerated using the original data—as opposed to *replicable* results, in which another group is able to draw similar conclusions from *new* data [11,12]. Below, we outline what constitutes code versus data, provide a how-to guide for code sharing, and conduct a survey of integrative biology journals to highlight how data versus code sharing requirements differ. Finally, we discuss which code best practices can be applied on inherited versus novel data sets.

## Journal policy survey

We performed a policy survey of prominent biology journals to determine the current state of publisher code-sharing requirements. We reviewed the author instructions and editorial policies for 100 of the most highly cited journals using data from the Scimago Journal & Country Rank portal [13]. Briefly, we started with a list of 10,433 journals in the life sciences, removed 7090 journals with less than 100 publications per year, and filtered out all journals that averaged less than 10 citations per article (see **Methods**), resulting in the final list of 100 journals.

We found that while the majority of journals now mandate the sharing of sequencing data, very few (10%) require authors to share analytical code. We found 10 journals that require code to be shared in some form: *Nucleic Acids Research* "requires all authors, where ethically possible, to provide access to all data and software code underlying the results presented in their article at submission" and, after acceptance, to publicly release "raw data, processed data, codes, scripts etc." [14]. *Radiology* also requires public deposition of "all computer code used for modeling and/or data analysis," as does *Science*, *Science Advances*, *Science Immunology*, and *Science Translational Medicine* [15]. The remaining four journals require sharing, though not via public deposition [16–19].

We further categorized journal policies by evaluating their requirements for sharing across three categories: sequencing data (or "large-scale datasets" such as proteomics), general datasets, and code. The majority of journals (55%) require public deposition of sequencing data, but only 11% require public deposition of all other data supporting a paper's results (**Figure 1**). A further 36% of journals require authors to share their raw (non-sequencing) data but allow it to be available "upon request." 10% of journals require the sharing of analytical code. For this analysis, we were searching for policies regarding analytical code, rather than software explicitly designed to be shared. Many journals have policies regarding the sharing or even peer review of code for articles describing new tools or novel algorithms, but we consider this a separate issue from sharing the details of computation that is incidental to the results. This latter category of computation is more likely to be done by biologists who are less familiar with programming best practices and community norms.

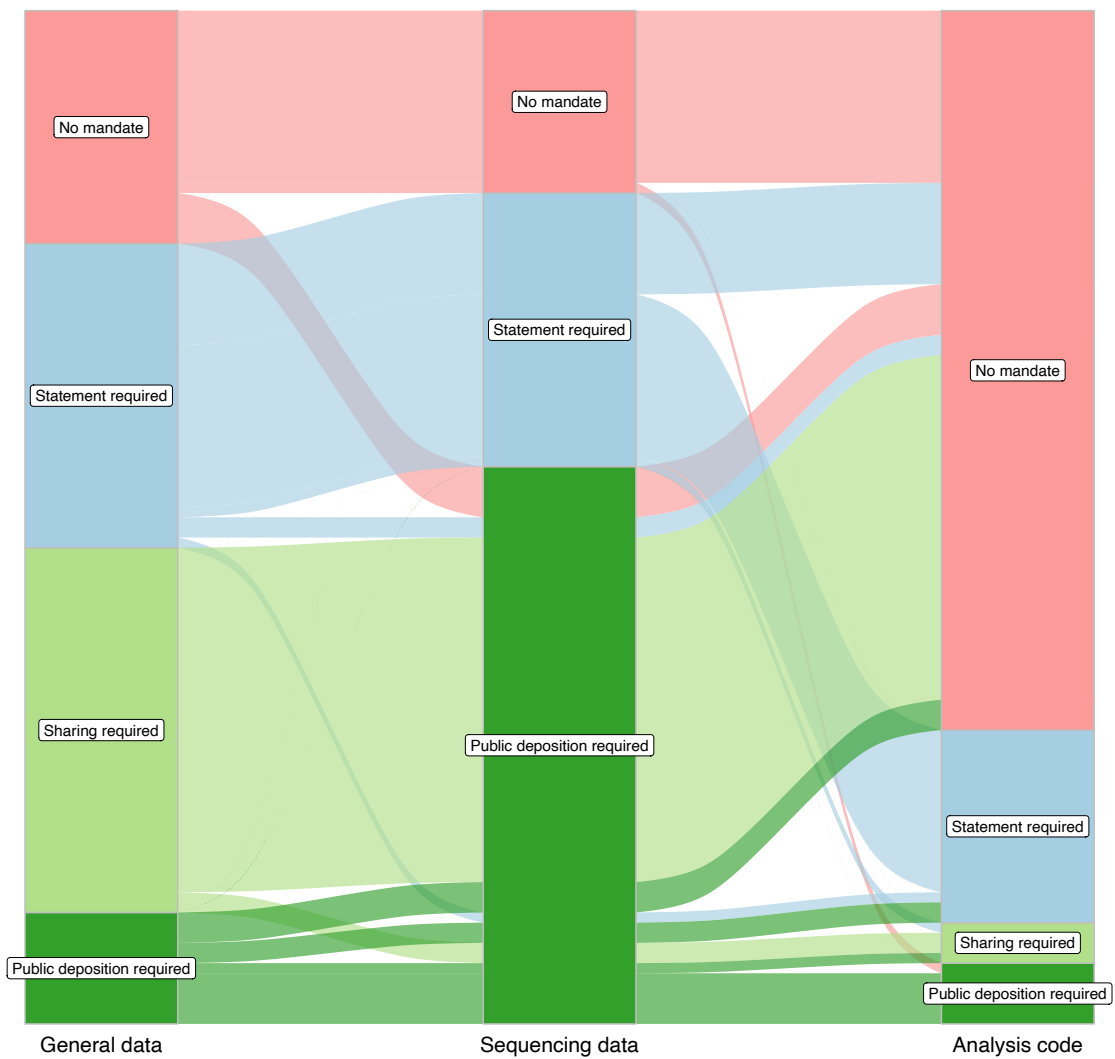

**Figure 1. Journal sharing policies.** An alluvial plot illustrating the relationship between journal requirements for the three types of data (General data, Sequencing data, and Analysis code). 'General data' includes policies for sharing general data supporting a paper's conclusions. 'Sequencing data' includes sequencing data and other large-scale datasets such as proteomics. (There were no journals in the light green "Sharing required" category for sequencing data.). 'Analysis code' includes code. Each of the 100 journals was placed into a single category for each of three data types, and the relative height of the colors represents the proportion of journals in each category for the given data type. The ribbons connecting each column are colored according to the "General data" category and illustrate the overlap between categories in neighboring columns: For example, the pink "No mandate" ribbon extending from the left column over to the "Public deposition required" category in the center column indicates that there are five journals with no mandate for sharing general data but that require public deposition of sequencing data. This pink ribbon then extends to the "No mandate" section in the third column, indicating all five journals also have no mandate for analysis code.

It is difficult to gauge whether journal-level code mandates are becoming more common: A 2013 study found 12 out of 170 surveyed journals (7.1%) required the sharing of code, though the journals were from a broader selection of fields including statistics and materials science [20]. And at least one biology journal has softened its position, at least on paper: *Science* announced in a 2011 editorial that they were "extending our data access requirement listed above to include computer codes involved in the creation or analysis of data" [15], and their policy has been quoted as saying, "All computer codes involved in the creation or analysis of data must also be available to any reader of *Science*" [20–23], at least as recently as 2016 [24]. By 2021, however, the policy was modified to its current form, which says code must be shared if it is "central to the findings being reported" [25,26]. Because of the 2011 editorial, we categorized *Science* (and its sibling journals listed above) as mandating public code deposition, though we suggest that an essay from 12 years ago should not be needed to clarify what a publisher requires. *Research*, a "Science Partner Journal" also published by the American Association for the Advancement of Science (AAAS), states this much more clearly, and mandates that readers have access to "all data and materials (including computer codes) necessary to understand, assess and extend the conclusions of the manuscript" [18].

Given the diversity of research environments and the changing technology landscape, it's not surprising that many policies are difficult to parse. For example, the *Journal of Medical Virology*, published by Wiley, states in their author instructions that the journal "expects that data supporting the results in the paper will be archived in an appropriate public repository" [27] However, the publisher's central policy page reveals that Wiley has multiple tiers of requirements, and that "expects data sharing" is a separate level from "mandates data sharing" [28]. Other publishers, such as PLOS, "expect" researchers to make "all author-generated code available without restrictions," but "may require" it for some papers [29].

Some publishers require sharing of some code, but not all. Cell Press, for example, requires sharing of only "original code reported in the paper," which suggests code actually discussed in-text, such as novel algorithm implementations, would be required, but general analysis code may not be; no examples are offered [16]. Another prominent publisher, BioMed Central (BMC), states that "Submission of a manuscript to a BMC journal implies that materials described in the manuscript, including all relevant raw data, will be freely available to any scientist wishing to use them for non-commercial purposes." However, the following sentence states, "BMC strongly encourages that all datasets on which the conclusions of the paper rely should be available to readers." Neither "strongly encourages" nor "implies" explicitly defines a mandate like the one described for sequencing data [30]. *Nature* and its sibling journals similarly state in bold text that "authors are required to make materials, data, code, and associated protocols promptly available to readers," but the later section on code availability clarifies that authors must only address whether they will be sharing "custom code or mathematical algorithm [*sic*] that is deemed central to the conclusions" [31]. The policy later notes, "Nature Portfolio journals consider it best practice to release custom computer code in a way that allows readers to repeat the published results," but stops short of a mandate.

There are also surprising gaps: For example, The editors of *Radiology* "request that all computer code used for modeling and/or data analysis be deposited in a publicly accessible repository upon publication," but their only data sharing guidance is in the section devoted to clinical trials [32]. *American Psychologist* more explicitly states something similar: Authors need only disclose whether they will share "the raw and/or processed data upon which study conclusions are based," but it is a journal requirement that authors must share all "computer code or syntax needed to reproduce analyses in an article" [17]. While this may work differently in practice, there are at least two journals with stated policies requiring authors to share analysis code but not the data it was actually analyzing. Interestingly, this inversion resembles results from Sharma et al. [33], in which 65% of surveyed papers published in 2021 shared code, but only 38% shared data.

## Methods

**Journal list.** We retrieved journal-level information from the Scimago Journal & Country Rank portal [13], which uses data from the Scopus database (https://www.scopus.com) to generate summary statistics for almost 28,000 academic publications. We downloaded the most recent data (2022 release) for all journals listed in the following categories: "Agricultural and Biological Sciences," "Biochemistry, Genetics and Molecular Biology," "Immunology and Microbiology," "Medicine," "Multidisciplinary," and "Neuroscience." After removing duplicates, this resulted in a list of 10,433 journals. We removed 7,090 journals that had fewer than 300 citable documents published over the last three years, and an additional 103 journals which averaged 80 or more references per document, to exclude review journals. We then filtered the list to include only journals that averaged at least 10 citations per document. This resulted in a list of 108 journals; after removing the remaining review journals (such as *Trends in Cancer* and *Current Opinion in Food Science*) that satisfied all other criteria but did not have applicable policies, we evaluated the remaining 100 journals manually.

Each journal's online policy documents and author instructions were reviewed independently by two authors. Each journal was placed in a single category reflecting its policies on sharing data and code; in order of openness: 1) no data or code sharing policy, 2) data sharing is encouraged, but not required, 3) an availability statement is required, but not sharing, and 4) data and/or code sharing is required. Where the two evaluators disagreed, the journal policies were re-evaluated to determine a final category. Journals who mandated any kind of sharing were then further evaluated at a more specific level that clarifies how the policies specifically apply to sequencing data, general data, and code.

# Recommendations

We can all share code—if not because someone else may use it in the future, then at least because *we* already did. The code used to analyze data, perform statistical tests and build visualizations is no less critical to a project than reagents used at the bench and should be disclosed for the same reasons: In addition to providing critical information about the conditions under which the study was performed, sharing this information enables others to more easily validate computationally derived conclusions. It also reduces the effort required to apply similar methods in new projects by providing an example to others which allows them to avoid issues you may have solved. A 2023 study across dozens of participants from 13 countries found broad support for publications that clearly state whether code was shared openly, with a persistent identifier (such as a DOI) and a clear license [34]. Sharing code also helps protect us against what Donoho et al. [35] called "the ubiquity of error" at the heart of the scientific method, which drives scientists to expend effort primarily "in recognizing and rooting out error": Even the most diligent scientists can miss a typo in a command or misunderstand parameters of a complex function in a package developed elsewhere. Sharing this code—essentially a fine-grained addendum to a methods section—can help bring these issues to the surface and help future researchers avoid them.

In this discussion, we use "code" as a shorthand for any of the documents interpreted by a computer to generate information used in your manuscript. That covers software you've developed to perform analyses (a new read alignment tool, for example), but it also includes commands used to perform statistical tests and the scripts used to generate figure panels. Most manuscripts don't come bundled with an entirely new software application, but many—especially those that include the analysis of genomic sequencing data—required code to be written for the generation of their results.

## Files to share

1. **New software applications.** If you wrote a whole new program to perform your work, it's essential to provide as much transparency as possible about how it works. The editorial staff at the journal should provide guidance about how to handle software that you don't intend to make open-source. Best practices for tool development have been well-covered elsewhere [36–41] and are outside the scope of this review, but if you've developed a tool that is useful but not the primary focus of your work, it could be helpful to submit a separate software paper or "application note" about your program to a computational journal [42].

2. **Scripts for data-cleaning and analysis.** Nothing is too mundane! These steps make it easier for others to reproduce your work and can clarify exactly what was done and, crucially, in what order: For example, it may seem silly to share the exact Python statement you used to perform a straightforward logistic regression, but the "LogisticRegression" function from the popular Python package scikit-learn defaults to using L2 regularization [43,44], while R's built-in "glm" function doesn't include similar penalties even as an option [45]—one of many potentially important details about model development [46] that may not be obvious but could be tracked down using the original code.

3. **Data visualization code.** Sharing the code you used to generate your figure panels can help people who want to visualize their data in a similar way and clarify finer points of the figure that have been omitted from the legend, intentionally or not. The code can also show exactly how data was filtered and modified before visualization. This could be useful to readers, who can now explore your results with "living figures" they can modify to look at different subsets of your data [47]; it may also serve as valuable documentation for your future reference.

4. **Parameters used to configure and launch command-line utilities.** Many computational biology tools such as Trimmomatic are executed from the Linux command line, with relevant parameters included directly in the command, specifying the location of input and output files, for example, or setting other configuration values such as thresholds or file formats. Ideally, a single script could be executed to run each command and perform your entire analysis process [48]. Even if that isn't how you executed these operations, however, including a file of ordered commands could be useful for those trying to evaluate minor implementation details, either for reproducing a paper's findings or for applying a similar process to their own data. (It's also worth noting that running commands using scripts is highly preferable to attempting to reconstruct these commands after the fact—see "Set yourself up for success" below for other techniques to keep in mind.)

5. **Pipeline specifications and configuration files.** Using workflow automation tools such as Snakemake [49] or Nextflow [50] can streamline your own bioinformatics work and make it easier to reproduce. Even in situations where your pipeline code is written to work only for your data or to run only on a specific computing cluster, the code can serve as valuable documentation of the process: which Docker containers were launched at each stage, which versions of software were used, and which options were passed to each tool can all be gleaned from this code by interested parties, even if they can't execute the exact code.

6. **A list of dependencies.** Providing a very specific list of all software dependencies in your pipeline may make a critical difference in how reliably your work can be reproduced: Python packages you import into your scripts, for example, may change dramatically between releases. For one, the popular Python library NumPy is used for many linear algebra operations, but the functionality of its matrix multiplication function was unintentionally changed when version 1.16.0 was released, then fixed with the release of version 1.16.6 nearly a year later [51,52]. There have been dozens of releases since, but a script that worked one way in 2019 may work very differently now—unless you record the version numbers. Placing a call to "sessionInfo()" within R scripts should print all package versions in the output of the script, for example, and in Python, running "pip freeze" (or its equivalent, if you're using a different tool such as conda for installing packages) will print out the installed versions of the packages installed in your environment. It would provide a more complete accounting to provide a Docker container with everything installed already, but even a list of library versions will cover most contingencies that don't involve low-level factors such as drivers and differences in hardware.

## How to prepare the code

First things first: Your code is good enough to share [53]! It may be messy and disorganized and cobbled together by self-taught coders writing just enough to get the job done, but you aren't the only person for whom that's true. If you trust the code enough that you've written a paper about its results, it's certainly worthy of sharing. Scientific code can be split into two broad categories: products that are intended to be reused by others (such as a new software package), and products that aren't. The preparation, packaging and archiving of code differs greatly between these two categories, and this paper deals mostly with the latter—code shared to demonstrate the computational approach to a single paper, but that others should not expect to work like a broadly applicable tool with a friendly user interface. The recommendations here describe considerations that are helpful, but if some are not practical to implement, it's important to note that whatever you can share is almost always better than sharing nothing.

The most important files to include are the source code files themselves—the files with extensions such as ".py" and ".R" and ".cpp" that were written by those performing the analyses. Byproducts of these scripts or their dependencies are not necessary to share—files such as those ending in ".pyc" and those created by the installation of packages are not necessary to include because they will be regenerated by the user doing their own installations. A simple exercise for distinguishing these files is to try re-running your code on a machine that was uninvolved in the original analysis—some files are regenerated by running the code on the new machine, some files can be downloaded instead, but the ones that you need to move manually are likely the ones most important to share.

Switching workstations can also help highlight aspects of your code that will get in the way of others trying to run it themselves. Removing things like passwords and API keys is critical, but other workstation-specific settings can also trip up others: hard-coded file paths will likely only work on a machine connected to that exact file system, even when pointing at common utilities. Storing strings like these in variables declared at the top of your script will provide clues to future users that this is a value they need to specify themselves, or an important variable that is used in multiple places. This is also why it's important to specify dependencies with enough detail that others can reproduce as much of your environment as possible. Other components of your pipeline may not be "dependencies" per se, but they can be critical to reproducing your work. Reference databases, for example, don't need to be shared with your code (assuming they are publicly available), but noting the version in your methods section is important—unsurprisingly, different reference databases can result in different results [54], even between minor versions of the same database [55]. Similarly, it is also helpful to note the operating system on which the code was run, particularly if you're using command-line utilities. Implementation details of tools such as "sed" can vary between platforms, and commands that work on a Linux machine may fail on macOS or even slightly different distributions of Linux—or worse, they may finish "successfully" and return different results.

You may also encounter situations where there are computational steps that do not have any executable scripts associated with them: perhaps the tool requires a commercial license, or there's no programmatic interface and the "point and click" approach is the only option. In situations like this, share whatever you can. It can be helpful to write out the computational steps both before and after the proprietary step and provide a bridge between sections of usable code by indicating where the difficult step is located. Providing users with the output of these tricky steps will enable them to skip over that step and pick up where you left off.

Finally, consider adding documentation to your repository; even a "README" text file providing a cursory description of how to follow the pipeline could make a big difference in how useful your files are to others, doubly so if you provide more detail about things like important parameters, installing tricky dependencies, and where to place input files.

Once you have pulled together all the documents you want to share, they must then be deposited. Repository services such as GitHub, GitLab and Sourceforge are reliable tools for collaboration and distribution [56], particularly for ongoing software projects, but many organizations such as university libraries [57,58] and groups like the Software Sustainability Institute [59] are skeptical of a reliance on the no-cost perpetual hosting of digital artifacts by commercial enterprises that haven't explicitly stated that as one of their goals. In contrast, repositories such as Zenodo are free, publicly funded projects with decades-long retention plans [60]. Depositing files in a repository that mints digital object identifiers (DOIs) improves their citability and makes links less likely to break over time by providing a canonical URL through the DOI resolution service available at doi.org. Deposits with these services are intended to be immutable artifacts, however, so it is preferable to wait until you have a "final" version before sharing there.

## Set yourself up for success

These recommendations were developed as guidance mainly to those who are preparing to share completed work. However, much of this will be easier—and more effective—if a project is set up with reproducibility in mind from the beginning. It can be frustrating to step back from investigation to set up peripheral systems like code repositories and Nextflow pipelines when an ad-hoc solution will move things forward more quickly. However, these shortcuts can backfire later when trying to repeat an analysis, recall who did which work, which subset of the data was used for a figure panel, or exactly which version of a program was used for an important computational step. Still, there is a balance to be found: Software developers looking at code from computational biology papers were shocked at its content, confusing structure and lack of documentation [61], but, as others have argued, optimizing process past a point of practical reproducibility may not be worthwhile [62], particularly when competing priorities leave researchers with few professional incentives to tackle the time-consuming work of sharing digital materials.

There are many guides for reproducible computational biology available online, both peer-reviewed and independently published [e.g. 9], covering everything from telling stories with computational notebooks [63,64] to standards and checklists that provide very specific examples [48,65]. Rather than duplicate this effort, here we review the most common recommendations and urge readers to investigate the implementation details in referenced papers and software documentation.

The first key to sharing your code is to use code in the first place: While many tools simplify their operation using a graphical user interface (GUI), this point-and-click approach can be difficult to document and even more difficult to replicate. Figuring out command-line tools and their parameterization can be a challenge, but running these operations a second time requires simply pasting the command into the terminal, rather than carefully following step-by-step instructions on which buttons to click in which order [66]. From here, code style and organization is a recurring theme in the literature. In short, consider approaching your code as if it were intended to be read, rather than executed [67]. Keep humans in mind: Give variables helpful names, rather than "foo" or "x," and leave plenty of comments to explain what different sections of code are doing and reasons for unconventional design decisions that users may be tempted to modify [66]. Use a consistent directory structure that makes it easy to distinguish raw data from intermediate files and final results [65,68,69]. When manipulating your data specifically for a visualization, save the version of the data that's actually displayed [70]: This will be a more convenient file for readers to evaluate, but it also means rebuilding your figure requires only loading that file, rather than loading the original dataset and performing all the processing steps again. Minimize manual intervention required to run your scripts, and ideally organize them in a way that allows users to run them from start to finish [48]. Provide a "README" file that, at minimum, walks a user through the intended execution of your code and documents key steps [71].

There is also guidance available for performing analysis. Importantly, emphatically avoid manual editing of intermediate files [9,69,70,72]—for example, if script A processes raw data into a table of gene expression levels, and script B summarizes this table by pathway, mental alarms should go off if there is a step after script A that requires a user to open the table and edit fields by hand to prepare the data for script B. Such quick fixes can be tempting, but they also add risk: A researcher could easily add a typo, corrupt a file in unexpected ways [73] or forget the step altogether.

Similarly, keep an eye out for opportunities to reduce coding errors by using custom functions to do repeated operations [9], and store important values in prominently commented variables, rather than hard-coding them somewhere deep in the script where a future user may not notice it. For example, if you write out a multi-step process that prepares data for a particular visualization, then find later that you need to perform the same steps for a different subset of data in a new figure panel, avoid copying that code and pasting it farther down in your analysis script—if you later find an error in this code or simply modify it to change a threshold or reorganize the output, it's easy to forget to scroll back down and change it in two (or more!) places. Even if you do remember, the background noise of keeping track of your various copy/paste maneuvers only builds over time. Repetitive code is one example of "code smell": Code that may work as intended, but that is suggestive of a larger design flaw that could cause hard-to-find bugs or make modification of the program more complicated [74]. Tools such as linters may help detect some of these issues, but eventually you will develop a "nose" for intuiting when you may be wandering down an ill-advised path.

There are many other recommendations that appear in the literature even without expanding your search beyond papers focused on computational biology: Version control and code review are frequent topics of discussion [9,66,69,75–77], as is "defensive programming" [67], or performing what may feel like excessive validation of the inputs and outputs of functions to make sure unexpected states are detected before they can cause problems. Automated testing is another practice that can make your code more robust, particularly for complicated procedures where a small change may have unintended effects elsewhere [78]. While modules such as "unittest" and frameworks like "pytest" can help you define and track granular verifications of specific functionality, it can be helpful to write a separate script, potentially with fake, easier to manage data, that you can run to validate that different sections of your code are returning expected values.

## Conclusion

Here, we have reviewed almost three dozen articles about reproducible research practices to summarize their recommendations for those seeking to comply with code-sharing requirements from journals or funders—or, more commonly, simply seeking to share their code because they are enthusiastic about participating in the open science ecosystem [30]. The factors involved in the sharing of data and code—and in open science issues more broadly—are complex, both for researchers and research subjects, particularly in relation to the equitable participation of marginalized communities [79–82]. If you've decided to share your code these recommendations will provide a starting point for the effort and connections to more detailed sources. Finally, we note that code sharing best practices should be taught early in the science curriculum alongside other Open Science approaches [83–87].

# Other information

## Funding

This study was supported by ASPIRE funds at the University of British Columbia- Okanagan (LG).

## Data and code availability

The detailed results of the journal policy survey and code used to generate Figure 1 are available at http://doi.org/10.5281/zenodo.10459940.

## Competing interests

The authors declare no conflicts of interest.

## Acknowledgements

We thank Casey S. Greene, Katja Della Libera, and Elizabeth Gibbons for helpful feedback.